\documentstyle[sprocl,epsf]{article}

\def\be{\begin{equation}}
\def\ee{\end{equation}}
\def\bea{\begin{eqnarray}}
\def\eea{\end{eqnarray}}

\begin{document}
\title{
{\small 
\begin{flushright}
CERN-TH/2000-092\\
March 2000
\end{flushright}}
RADIATIVE HIGGS-SECTOR CP VIOLATION IN THE MSSM}

\author{APOSTOLOS PILAFTSIS\footnote[1]{To appear in the Proc.\ 
of PASCOS '99, Lake Tahoe, California, December 10--16, 1999} }

\address{Theory Division, CERN, CH-1211 Geneva 23, Switzerland}

\maketitle\abstracts{  We    briefly  review the      phenomenological
implications of the  minimal supersymmetric standard model (MSSM) with
explicit  radiative breaking of CP  invariance in the Higgs sector for
the LEP2 and Tevatron    colliders. }

It has recently been shown \cite{AP} that the tree-level CP invariance
of the  MSSM Higgs  potential can sizeably  be broken at  the one-loop
level  by large  soft CP-violating  trilinear couplings  of  the Higgs
bosons  to   stop  and   sbottom  squarks.   Several   recent  studies
\cite{PW,Demir,CDL,CEPW,KW}  have  been  devoted  to analyze  in  more
detail  the  effective  Higgs  potential  of the  MSSM  with  explicit
radiative breaking of CP invariance.  We shall briefly review the main
phenomenological implications of this  rather rich and very predictive
theoretical framework of the MSSM for the LEP2 and Tevatron colliders.

In the $\overline{\rm MS}$ scheme, the one-loop CP-violating effective
potential of the MSSM is given by
\begin{equation}
  \label{LVeff}
-{\cal L}_V\ =\ -{\cal L}^0_V\, +\,
\frac{3}{32\pi^2}\, \sum\limits_{q = t,b} \bigg[\,\sum\limits_{i = 1,2} 
\widetilde{m}^4_{q_i}\, \bigg( \ln \frac{\widetilde{m}^2_{q_i}}{Q^2}\, -\,
\frac{3}{2}\,\bigg)\, -\, 2\bar{m}^4_q\, 
\bigg( \ln \frac{\bar{m}^2_q}{Q^2}\, -\, \frac{3}{2}\,\bigg)\, \bigg]\, ,
\end{equation}
where ${\cal  L}^0_V$ is the  tree-level Lagrangian of the  MSSM Higgs
potential,   and  $\bar{m}_q$   and   $\widetilde{m}_{q_i}$  are   the
field-dependent quark and squark masses of the third generation. Here,
we  adopt the  notation of  Refs.\ [1,2,5].   The minimization  of the
CP-violating   effective   potential   differs   from  that   in   the
CP-conserving  case by the  presence of  a non-trivial  CP-odd tadpole
condition  \cite{AP}  of  the  would-be  CP-odd scalar  $a$.   In  the
$\overline{\rm  MS}$  scheme,   the  CP-odd  tadpole  condition  reads
\cite{AP,note}
\begin{eqnarray}
\label{TA}
T_A\ =\ -v\, {\rm Im} (m^2_{12} e^{i\xi}) &=& -\, \frac{3}{16\pi^2}\, 
\sum_{q=t,b}\, \frac{s_{2q}}{s_\beta}\ {\rm Im}\, h^q_1\ \Delta
m^2_{\tilde{q}}\ B^{\rm fin}_0 (0,m^2_{\tilde{q}_1},m^2_{\tilde{q}_2})\, ,
\end{eqnarray} 
where $t_\beta =  s_\beta/c_\beta =  v_2/v_1$, $s_{2q}=  2\sin\theta_q
\cos\theta_q$,  $\Delta    m^2_{\tilde{q}}    =     m^2_{\tilde{q}_2}-
m^2_{\tilde{q}_1}$, $h^t_1 =  -m_t \mu^*  e^{-i\delta_t}/(s_\beta v)$,
$h^b_1 = m_b A_b^*  e^{i\delta_b}/(c_\beta v)$, $\delta_{t(b)}  = {\rm
arg}\, [A_{t (b)} -\mu^* t_\beta (1/t_\beta)]$ and
\begin{equation}
   \label{B00} 
B^{\rm fin}_0(0,m^2_1,m^2_2)\ =\ -\,
\ln\bigg(\frac{m_1 m_2}{Q^2}\bigg)\, +\, 1 +\, \frac{m^2_1 +
m^2_2}{m^2_1 - m^2_2}\, \ln\bigg(\frac{m_2}{m_1}\bigg)\, .
\end{equation}
Furthermore, $\mu$  and $m^2_{12}$ are respectively the  SUSY and soft
SUSY-breaking Higgs-mixing terms, $A_{t,b}$ are the soft SUSY-breaking
Yukawa couplings, and $\theta_q$ is  the mixing angle between the mass
and weak squark eigenstates.

To  one-loop order,  CP violation  is introduced  into the  MSSM Higgs
potential  through the  complex parameters  $\mu$ and  $A_{t,b}$.  The
CP-violating  terms  are   proportional  to  the  rephasing  invariant
combination: ${\rm Im}\, (m^{2*}_{12}  A_{t,b} \mu)$. {}From this last
expression,  it is  clear that  only the  relative phase  between $\mu
A_{t,b}$  and  $m^2_{12}$  plays  a  role.  Therefore,  a  good  phase
convention is  to define  $m^2_{12}$ to be  real.  This can  always be
achieved  by a  global  U(1) rotation.\cite{AP,PW}  As  a result,  the
relative phase  $\xi$ between the two Higgs  vacuum expectation values
$v_1$ and $v_2$  vanishes at the tree level.  The  phase choice $\xi =
0$  can  be preserved  order  by order  in  perturbation  theory by  a
corresponding  choice of  the counter-term  of ${\rm  Im}\, m^2_{12}$,
exactly as  is given in  Eq.\ (\ref{TA}).\cite{AP,CEPW} In  fact, this
perturbative resetting of the phase $\xi$ to zero is equivalent to the
general  requirement that  within  the effective-potential  formalism,
$v_1$, ${\rm  Re}\, v_2 = |v_2|  \cos\xi$ and ${\rm Im}\,  v_2 = |v_2|
\sin\xi$ do not receive finite radiative shifts in higher orders other
than  those due  to Higgs  wave-function renormalization.   The latter
approach is also consistent with the one followed in the CP-conserving
studies.\cite{Mh}

It is known that the MSSM faces the difficulty of explaining naturally
the apparent absence of electric  dipole moments (EDMs) of the neutron
and  electron.\cite{EFN,DDLPD,PN,IN,EF,CKP,APino}  Several suggestions
have  been made  to suppress  the SUSY  contributions to  electron and
neutron EDMs, at  a level just below their  present experimental upper
limits.   Apart  from  the  obvious  choice  of  suppressing  the  new
CP-violating phases of the  theory to the $10^{-3}$ level,\cite{EFN} a
more phenomenologically appealing possibility is to make the first two
generations of scalar fermions as  heavy as few TeV,\cite{PN} but keep
the soft-breaking  mass parameters of the  third generation relatively
small, e.g.\  0.5--0.7 TeV.  An interesting alternative  is to arrange
for partial cancellations among the different EDM contributions either
at   the  short-distance   level\cite{IN}   or  the   non-perturbative
long-distance one.\cite{EF}

In  addition  to the  one-loop  EDM  contributions  of the  first  two
generations, one  may have to  worry that third-generation  squarks do
not induce observable effects on the electron and neutron EDMs through
the three-gluon operator  \cite{DDLPD}, through the effective coupling
of  the `CP-odd'  Higgs  boson  to the  gauge  bosons \cite{CKP},  and
through  two-loop gaugino/higgsino-mediated  EDM  graphs~\cite{APino}. 
For low-$t_\beta$ scenarios, the  two-loop EDM contributions are found
to  be  of  the  order  of the  experimental  upper  bounds.\cite{CKP}
Therefore, it not very difficult to arrange the different two-loop EDM
terms  to  partially cancel  one  another,\cite{APino}  and so  reduce
significantly their total size.

An immediate consequence of CP violation in the Higgs potential of the
MSSM  is the  presence of  mixing-mass terms  between the  CP-even and
CP-odd Higgs fields.\cite{AP} In the weak basis $(\phi_1, \phi_2, a)$,
the neutral Higgs-boson mass matrix ${\cal M}^2_N$ takes on the form
\begin{equation}
  \label{NHiggs}
{\cal M}^2_N \ =\ 
\left[ \begin{array}{cc} {\cal M}^2_S  & {\cal M}^2_{SP} \\
        ({\cal M}^2_{SP})^T &  {\cal M}^2_P \end{array} \right]\, ,
\end{equation}
where  ${\cal M}^2_S$  and ${\cal  M}^2_P$ describe  the CP-conserving
transitions between  scalar and pseudoscalar  particles, respectively,
whereas  ${\cal M}^2_{SP}$ describes  CP-violating scalar-pseudoscalar
transitions.    The   characteristic   size  of   these   CP-violating
off-diagonal  terms in  the Higgs-boson  mass matrix  was found  to be
\cite{AP,PW}
\begin{eqnarray}
  \label{MSP}
M^2_{SP} & \simeq & {\cal O} \left( \frac{ m_t^4}{v^2} 
\frac{|\mu| |A_t|}{32 \pi^2M_{\rm SUSY}^2} \right) \sin \phi_{\rm CP} 
\nonumber\\
&&\times\, \left(6,\ \frac{|A_t|^2}{M_{\rm SUSY}^2}\, ,\ \frac{|\mu|^2}
{\tan\beta\, M_{\rm SUSY}^2}\,,\ \frac{\sin 2\phi_{\rm CP}}{\sin
  \phi_{\rm CP}}\, \frac{|\mu||A_t|}{M_{\rm SUSY}^2} \right),
\end{eqnarray}  
where the last bracket summarizes the  relative sizes of the different
contributions, and $\phi_{\rm  CP} =  {\rm  arg}(A_t \mu)$. As  can be
seen from  Eq.\   (\ref{MSP}),  the CP-violating  effects   can become
substantial if $|\mu|$ and $|A_t|$ are larger  than the average of the
stop masses, denoted as $M_{\rm SUSY}$.  For example, the off-diagonal
terms  of the  neutral Higgs-mass matrix  may be  of order $(100\ {\rm
GeV})^2$,  for   $|\mu| \simeq   |A_t|  \stackrel{<}{{}_\sim} 3 M_{\rm
SUSY}$, and $\phi_{\rm  CP} \simeq 90^\circ$.  

\begin{figure}
   \leavevmode
 \begin{center}
   \epsfxsize=12.0cm
    \epsffile[0 0 539 652]{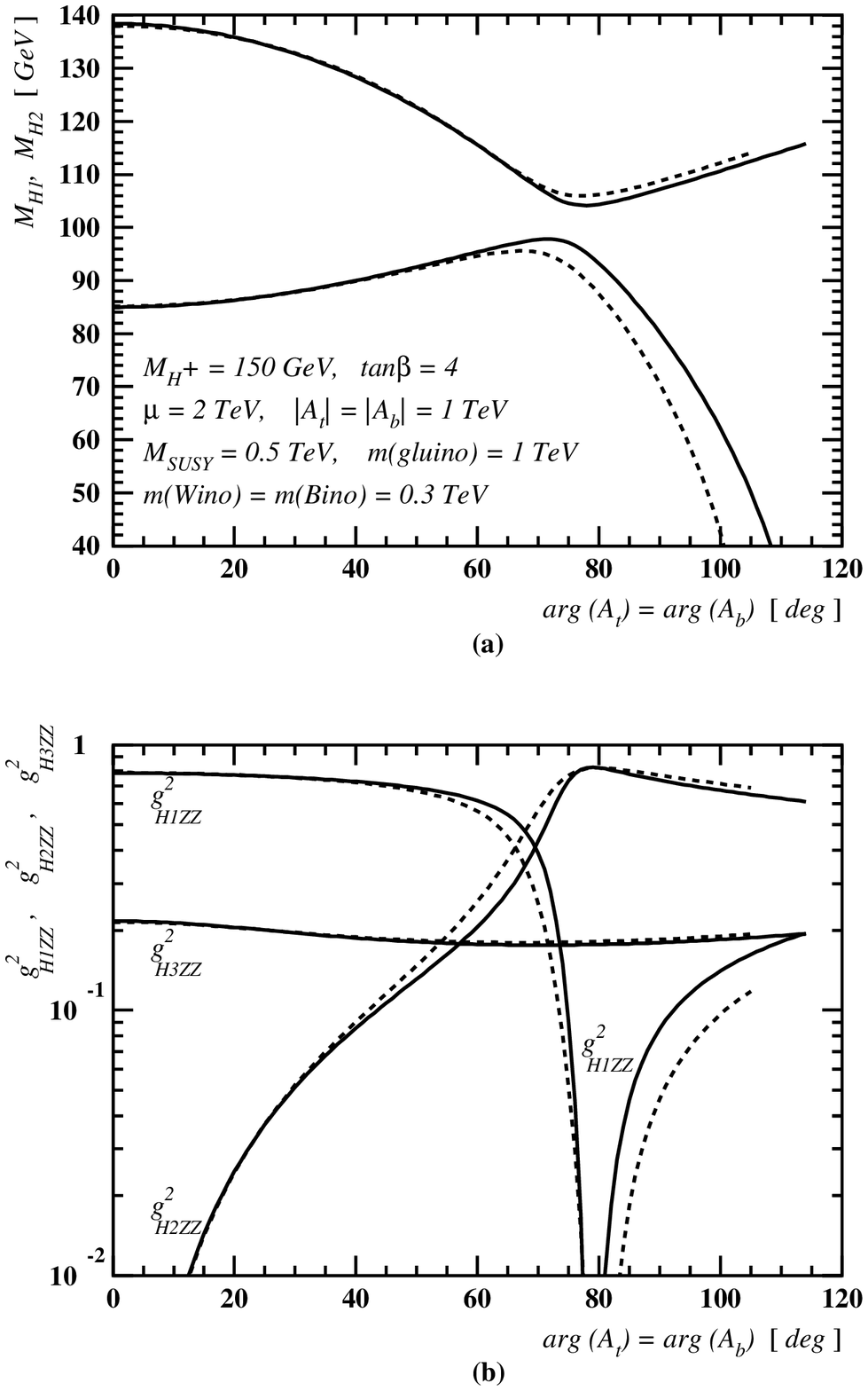}
 \end{center}
 \vspace{-0.5cm} 
\caption{Numerical estimates~{}$^5$  of (a) 
  $M_{H_1}\le  M_{H_2}$ and  (b) $g^2_{H_iZZ}$  and as  a  function of
  ${\rm arg}\, (A_t)$.}
\label{fig:pascos1}
\end{figure}
The main  effect of Higgs-sector  CP violation  is the modification of
the couplings  of  the Higgs  bosons to fermions  and the  $W$ and $Z$
bosons,  i.e.\ $ffH_i$, $WWH_i$,  $ZZH_i$  and $ZH_iH_j$. The modified
effective Lagrangians are given by
\begin{eqnarray}
  \label{Hiff}
{\cal L}_{H\bar{f}f} &=& -\, \sum_{i=1}^3 H_i\,
\Big[\, \frac{g_w m_{d}}{2M_W c_\beta}\, \bar{d}\,( O_{1i}\, -\,
is_\beta O_{3i}\gamma_5 )\, d\nonumber\\
&& +\, \frac{g_w m_u}{2M_W s_\beta}\, \bar{u}\,( O_{2i}\, -\,
ic_\beta O_{3i}\gamma_5 )\, u\, \Big]\, ,\\
  \label{HVV}
{\cal L}_{HVV} &=&  g_w M_W\, \sum_{i=1}^3\, ( c_\beta O_{1i}\, +\, s_\beta
O_{2i} )\, \Big(\, H_i W^+_\mu W^{-,\mu}\nonumber\\
&&+\ \frac{1}{2c^2_w}\, H_i Z_\mu Z^\mu\, \Big)\, ,\\
  \label{HHZ}
{\cal L}_{HHZ} &=& \frac{g_w}{4c_w}\, \sum_{i,j=1}^3\,
\Big[\, O_{3i}\, ( c_\beta O_{2j}\, -\, s_\beta
O_{1j} )\, -\, O_{3j}\, ( c_\beta O_{2i}\, -\, s_\beta
O_{1i} )\, \Big]\, \nonumber\\
&&\times\, Z^\mu\, ( H_i\, \!\!
\stackrel{\leftrightarrow}{\vspace{2pt}\partial}_{\!\mu} H_j )\, ,
\end{eqnarray}
where $c_w =     M_W/M_Z$,    $\stackrel{\leftrightarrow}{\vspace{2pt}
  \partial}_{\!   \mu}\   \equiv\  \stackrel{\rightarrow}{\vspace{2pt}
  \partial}_{\!    \mu}     -       \stackrel{\leftarrow}{\vspace{2pt}
  \partial}_{\! \mu}$, and $O$ is the orthogonal transformation matrix
  relating        the    weak    with       the    mass    Higgs-boson
  eigenstates.\cite{PW,CEPW}
 
    Let us  now discuss  a representative  example  demonstrating the
phenomenological consequences of Higgs-sector CP violation on the LEP2
and  Tevatron  colliders.   We  consider  an  intermediate  value  for
$\tan\beta = 4$, and a relatively light charged Higgs boson $M_{H^+} =
150$ GeV, with $M_{\rm SUSY} = 0.5$  TeV, $A_t = A_b = 1$ TeV and $\mu
= 2$ TeV.  In Fig.\  \ref{fig:pascos1}, we then find regions for which
the lightest Higgs-boson mass $M_{H_1}$  is as small as 60--70 GeV for
${\rm  arg}   (A_t)  \approx  90^\circ$,  and   the  $H_1ZZ$  coupling
$g_{H_1ZZ}$, which is normalized to  its SM value, is small enough for
the  $H_1$ boson  to  escape detection  at  the latest  LEP2 run  with
$\sqrt{s}  = 202$  GeV.\cite{CEPW} Moreover,  the $H_2$  boson  is too
heavy to be detected through the $H_2ZZ$ channel.  In addition, either
the  coupling $H_1H_2Z$, $g_{H_1H_2Z}  = g_{H_3ZZ}$,  is too  small or
$H_2$  is  too  heavy  to  allow  Higgs  detection  in  the  $H_1H_2Z$
channel.\cite{CEPW}  An upgraded  Tevatron machine  has  the potential
capabilities to  close most of such experimentally  open windows.  The
results of  the recent complete RG  analysis of Ref.\ [5]  are in good
qualitative agreement with earlier studies.\cite{PW,CDL}

In  conclusion,  the  MSSM  with  explicit radiative  breaking  of  CP
invariance in the  Higgs sector \cite{AP} constitutes a  very rich and
predictive  theoretical framework,  with  interesting consequences  on
collider experiments,\cite{AP,Demir,PW,CDL,CEPW,KW,GGK} CP asymmetries
in     $B$-meson     decays,     \cite{Bmeson}     and     electroweak
baryogenesis.\cite{EW}

I  wish to  thank Marcela  Carena, Darwin  Chang, John  Ellis, Wai-Yee
Keung and Carlos Wagner for collaboration.



\begin{thebibliography}{9}
\bibitem{note} See Eq.\ (27) in the second reference in [1] [cf.\
Eqs.\ (4.8) and (4.9) in the first reference of [1]]. Note that 
Eq.\ (2) fully agrees with subsequent calculations in [2--6].
\end{thebibliography}
\end{document}